\documentclass[aps,prl,twocolumn,superscriptaddress,showpacs]{revtex4}
\usepackage{graphicx}
\usepackage{dcolumn}

\usepackage{bm}
\usepackage{color}
\usepackage[colorlinks]{hyperref}
\input{epsf}
\usepackage{verbatim}
\usepackage[sort&compress]{natbib}
\begin{document}


\title{Nonlinear growth with the microwave intensity in radiation-induced magnetoresistance oscillations}

\author{R. G. Mani}
\affiliation{Department of Physics and Astronomy, Georgia State
University, Atlanta, GA 30303 U.S.A. }
\author{C. Gerl}
\affiliation{Institut f\"{u}r Experimentelle und Angewandte
Physik, Universit\"{a}t Regensburg, 93053 Regensburg, Germany}
\author{S. Schmult}
\affiliation{Institut f\"{u}r Experimentelle und Angewandte
Physik, Universit\"{a}t Regensburg, 93053 Regensburg, Germany}
\affiliation{ Max-Planck-Institut f\"{u}r Festk\"{o}rperforschung,
70569 Stuttgart, Germany}
\author{W. Wegscheider}
\affiliation{Institut f\"{u}r Experimentelle und Angewandte
Physik, Universit\"{a}t Regensburg, 93053 Regensburg, Germany}
\affiliation{Laboratorium f\"{u}r Festk\"{o}rperphysik,
ETH-Z\"{u}rich, 8093 Z\"{u}rich, Switzerland}
\author{V. Umansky} \affiliation{Braun Center for
Submicron Research, Weizmann Institute, Rehovot 76100, Israel}

%
%
%
%
\date{\today}
\begin{abstract}
We report the observation of inverse-magnetic-field-periodic,
radiation-induced magnetoresistance oscillations in GaAs/AlGaAs
heterostructures prepared in W. Wegscheider's group, compare their
characteristics with similar oscillations in V. Umansky's
material, and describe the lineshape variation vs. the radiation
power, $P$, in the two systems. We find that the radiation-induced
oscillatory $\Delta R_{xx}$, in both materials, can be described
by $\Delta R_{xx} = -A exp(-\lambda/B)sin(2 \pi F/B)$, where $A$
is the amplitude, $\lambda$ is the damping parameter, and $F$ is
the oscillation frequency. Both $\lambda$ and $F$ turn out to be
insensitive to $P$. On the other hand, $A$ grows nonlinearly with
$P$.
\end{abstract}
%
%
\maketitle
\section{introduction}
Vanishing electrical resistance is an interesting characteristic,
and superconductivity is a well known example of this
phenomenon.\cite{1} Another example occurs in the quantum Hall
two-dimensional electronic system (2DES), where the application of
a magnetic field induces zero-resistance states in conjunction
with quantized Hall resistances at low temperatures.\cite{2,3} A
third example occurs in the microwave and terahertz irradiated
2DES where the radiation induces  vanishing resistance states
without concurrent Hall quantization.\cite{4,5} This latter
example has shown the possibility of photo-exciting into
zero-resistance states in a condensed matter system.

Photo-excited transport in the 2DES has become a topic of
experimental and theoretical interest in the recent past.\cite{4,
5, 6, 7, 8, 9, 10, 11, 12, 13, 14, 15, 16, 17, 18, 19, 20, 21, 22,
23, 24, 25, 26, 27, 28, 29, 30, 31, 32, 33, 34, 35, 36, 37, 38,
39, 40, 41} Periodic in $B^{-1}$ radiation-induced
magnetoresistance oscillations, which lead into the
radiation-induced zero-resistance states, are now understood to be
a consequence of radiation-frequency ($f$) and magnetic field
($B$) dependent, scattering at impurities \cite{ 24, 26, 27}
and/or a change in the distribution function.\cite{6,30,36} And,
vanishing resistance at the oscillatory minima is asserted to be
an outcome of negative resistance instability and current domain
formation.\cite{25, 32, 39} Although there has been much progress,
there remain many aspects that could be better understood from
both the experimental and theoretical perspectives. Here, some
open problems include understanding (a) the observed activated
temperature dependence at the resistance minima, (b) the role of
the potential landscape in influencing the magnitude of the
observed oscillatory effect, and (c) the nature of the overlap
with quantum Hall effect.\cite{23}

A further topic of interest is to examine, in detail, the growth
of the oscillatory effect vs. the radiation intensity, $P$. So far
as theoretical results regarding this aspect are concerned, a
number of works have numerically evaluated the radiation-induced
magnetoresistance oscillations for several $P$ and graphically
presented the results.\cite{24, 31, 33} In contrast, Dmitriev and
co-workers,\cite{30} have made the prediction that, in the linear
response regime, the correction to the dark dc conductivity is
linear in $P$, see eqn. 16, ref. \cite{30}.

A comparison of experiment with theory, so far as the
$P$-dependence is concerned, could help to identify the relative
importance of the invoked-mechanisms in the above mentioned
theories. Further, the novel proposal that these radiation-induced
phenomena could constitute a new example of complex emergence,
where self-organization can arise from a number of basic
interactions with remarkable collective aspects,\cite{34} has also
identified the need to better understand the $P$ dependence.
Finally, as a related issue, it is also of interest to compare the
growth of the radiation-induced resistance oscillations with $P$
in materials prepared in different laboratories in order to
determine whether wafer preparation practices, which influence the
scattering landscape within the electronic system and impact the
lifetimes / observed characteristics, lead to a perceptible
difference in the measurements.

Thus, we examine the growth of the radiation-induced
magneto-resistance oscillations with $P$ in GaAs/AlGaAs devices.
We also compare the radiation-induced transport in devices
fabricated from MBE material grown by Wegscheider (W), with
results from similar material prepared by Umansky (U). We find
that the radiation-induced oscillatory diagonal resistance,
$\Delta R_{xx}$, in both materials can be described by $\Delta
R_{xx} = -A exp(-\lambda/B)sin(2 \pi F/B)$, where $A$ is the
amplitude, $\lambda$ is the damping parameter, and $F$ is the
$f$-dependent magneto-resistance oscillation frequency. Both
$\lambda$ and $F$ turn out to be insensitive to $P$ and the
temperature, $T$. On the other hand, $A$ grows nonlinearly with
$P$, and the non-linearity depends on the $T$. Such growth of $A$
with $P$ and $T$ has not been predicted.

\section{Experiment}

Low frequency lock-in based electrical measurements were carried
out at $T \leq 1.5 K$ with the samples mounted inside a solenoidal
magnet near the open end of a waveguide that is also closed at the
other end.\cite{4,13} The samples were immersed in pumped
liquid-helium, and the temperature was determined with the aid of
calibrated resistance thermometers and the vapor-pressure
thermometry. Microwaves were conveyed via the waveguide, and the
intensity was set at the source, as indicated. Hardware between
source and sample introduces intensity attenuation. A standing
wave pattern also produces an axial intensity variation within the
waveguide, with a possible further intensity reduction at the
sample. Since the $B$-field readout of a superconducting magnet
power supply is not always reliable at low $B$, the $B$-field
could be characterized in-situ by performing Electron Spin
Resonance (ESR) of DiPhenyl-Picryl-Hydrazal (DPPH). The DPPH-ESR
identifies $B$ via the relation $B_{ESR} = f_{ESR}/[28.043
GHz/T$].\cite{8} The W-GaAs/AlGaAs single heterostructures were
nominally characterized by an electron density, $n = 2.4 \times
10^{11} cm^{-2}$ and a mobility of $\mu = 10^{7} cm^{2}/Vs$. The
U-material was roughly comparable, albeit with a slightly higher
$n$, $n = 3 \times 10^{11} cm^{-2}$. The 2DES's were prepared, as
is usual, using the Persistent Photo-Conductivity (PPC) effect,
with a brief illumination by a LED. Since the realized
low-temperature $\mu$ and $n$ depends upon the cool-down and
illumination procedure via the PPC effect, a given specimen can
show associated preparation dependent changes in the
radiation-induced characteristics. Results are reported here for
two sets of measurements, labeled U1 and U2, on a U-specimen, and
measurements, labeled W1, W2, and W3 on a set of three
W-specimens.
\section{Results}
\begin{figure}[t]
\begin{center}
\leavevmode \epsfxsize=3 in
 \epsfbox {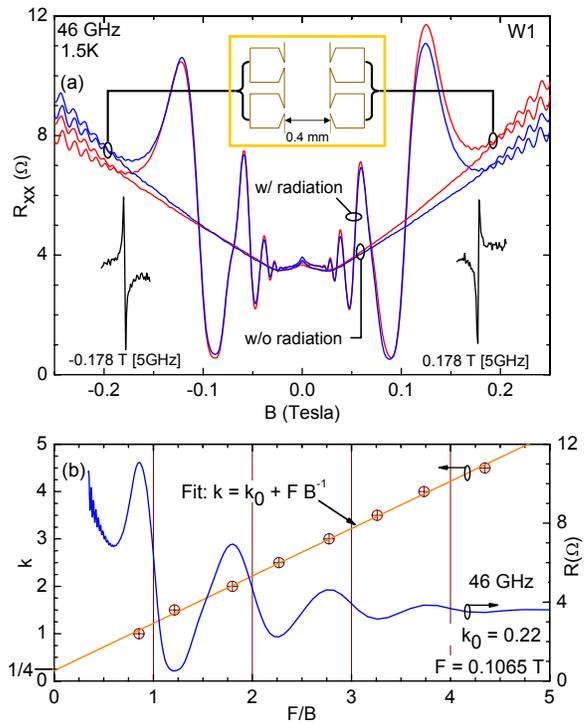}
\end{center}
\caption{(a) Inset: A sketch of the Hall bar sample. The main
panel exhibits $R_{xx}$ obtained in the dark (w/o radiation) and
under photoexcitation (w/radiation), both on the left (blue
traces) and right (red traces) sides of the device W1. $B$-field
markers based on ESR of DPPH at $f_{ESR}=5GHz$ occur at $\pm 0.178
T$. (b) $R_{xx}$, shown on the right ordinate, has been plotted
vs. $B^{-1}$ in order to exhibit the $B^{-1}$ periodicity of the
oscillations. The oscillatory maxima (minima) have been assigned
with integers (half-integers), see the left ordinate. The plot of
$k$ (left ordinate) vs. $B^{-1}$ has been subjected to a linear
fit (orange curve), i.e., $k = k_{0} + F B^{-1}$. Here, $k_{0}$
represents the phase, and $F$ is the frequency of the
radiation-induced magneto-resistance oscillations. Thus, the plot
confirms a "1/4-cycle" phase shift in the W-specimen. The abscissa
has been scaled to $F/B$.} \label{mani01fig}
\end{figure}

Figure 1(a) exhibits the data for a $0.4 mm$ wide Hall bar (W1)
fabricated from the W-material. The figure shows the $R_{xx}=
V_{xx}/I$ measured both in the dark (w/o radiation) and under $46
GHz$ photo-excitation (w/ radiation). Also shown are $B$
calibration markers obtained from the ESR of DPPH at $f_{ESR}=5
GHz$. The figure shows large amplitude radiation-induced
magneto-resistance oscillations, as the $R_{xx}$ is reduced to
roughly $10\%$ of the dark value at deepest resistance minimum.
Further, the w/o radiation $R_{xx}$ trace intersects the
w/radiation $R_{xx}$ traces at the nodes of the
oscillations.\cite{4,24} Similar results are obtained on both the
left- and right- sides of the device. The small asymmetry observed
under $B$-reversal is attributed to an admixture between the
diagonal and off-diagonal (Hall) signals.
\begin{figure}[t]
\begin{center}
\leavevmode \epsfxsize=3 in
 \epsfbox {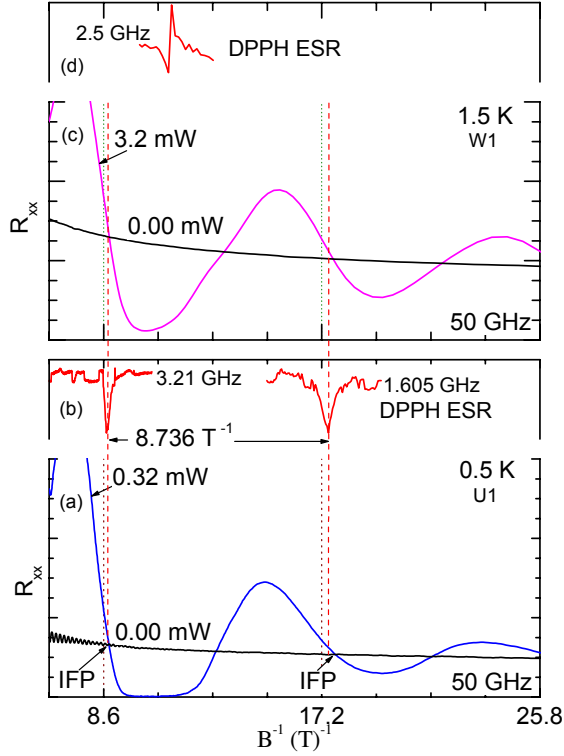}
\end{center}
\caption{(a) The  $50 GHz$ photo-excited and dark  $R_{xx}$ has
been plotted vs. $B^{-1}$ for U1. (b) The Integral Fixed Points
(IFP) DPPH-ESR $B$-field markers have been shown here for the
measurement shown in (a). (c) The $50 GHz$ photo-excited and dark
$R_{xx}$ for W1. (d) A DPPH ESR $B$ calibration marker for the
measurement in (c). This figure indicates good agreement between
the radiation-induced oscillatory magneto-resistance for different
MBE material. } \label{mani02fig}
\end{figure}

\begin{figure}[t]
\begin{center}
\leavevmode \epsfxsize=3 in
 \epsfbox {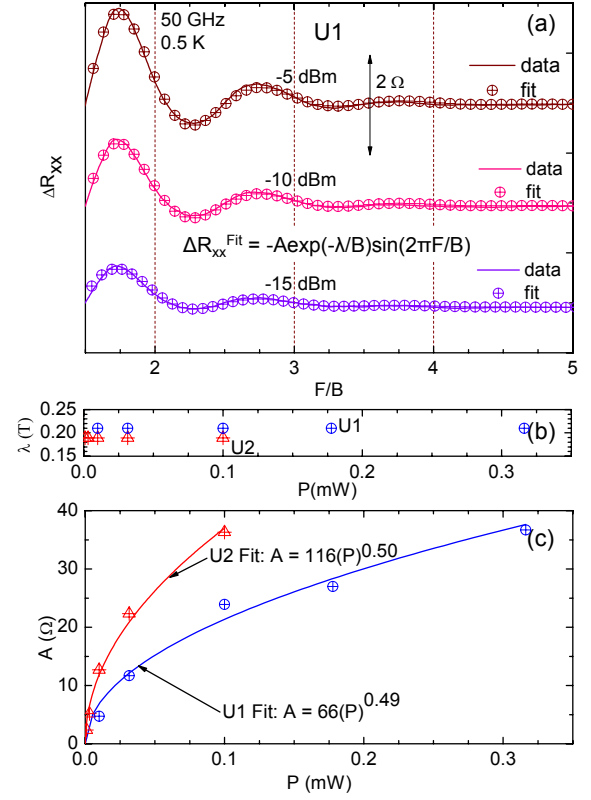}
\end{center}
\caption{(a) For U1, $\Delta R_{xx}$ is exhibited at $f=50GHz$
with $P$ as parameter. Also shown are fits to an exponentially
damped sinusoid, see text. (b) The damping constant, $\lambda$, is
plotted vs. $P$ for U1 and U2. (c) The lineshape amplitude, $A$,
is plotted vs. $P$ for U1 and U2. Also shown here are fits, $A =
A_{0} P^{\alpha}$, which suggest $\alpha = 0.49$ and $\alpha =
0.50$ for U1 and U2, respectively.} \label{mani03fig}
\end{figure}

In Fig. 1(b), $R_{xx}$ on the right ordinate has been plotted vs.
$B^{-1}$ in order to exhibit the $B^{-1}$ periodicity of the
oscillations. The oscillatory maxima (minima) were then assigned
with integers (half-integers), see the left ordinate, beginning
with $k = 1$ for the $R_{xx}$ maximum at the lowest $F/B$. The
plot of $k$ (left ordinate) vs. $B^{-1}$ has been subjected to a
linear fit, i.e., $k = k_{0} + F B^{-1}$, which is represented by
the orange curve in Fig. 1(b). Here, $k_{0}$ represents the phase,
and $F$ is the frequency of the radiation-induced
magneto-resistance oscillations. Thus, the fit extracted $k_{0} =
0.22$, see Fig. 1(b), confirms a "1/4-cycle phase shift".\cite{8}
In addition, $F = 0.1065 T$ at $f=46 GHz$, suggests that $ m^{*}/m
= e F/(2 \pi mf) = 0.065$, slightly lower than the standard value,
$m^{*}/m = 0.067$, for GaAs/AlGaAs system.\cite{8} In Fig. 1(b),
the abscissa has been scaled to $F/B$, using the $F$ indicated
above.

\begin{figure}[t]
\begin{center}
\leavevmode \epsfxsize=3 in
 \epsfbox {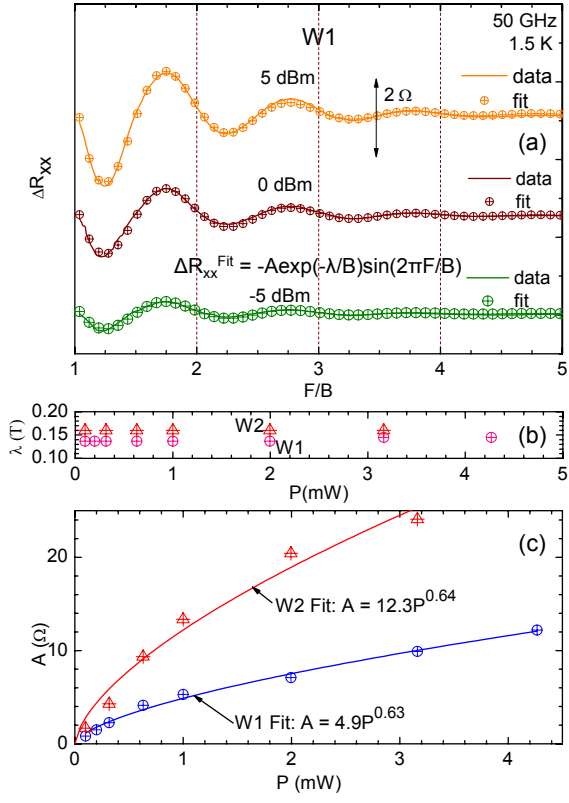}
\end{center}
\caption{(a) For W1, $\Delta R_{xx}$ is exhibited at $f=50GHz$.
Also shown are fits to an exponentially damped sinusoid. (b)
$\lambda$ is plotted vs. $P$ for W1 and W2. (c) The lineshape
amplitude, $A$, is plotted vs. $P$ for W1 and W2. Also shown are
fits, $A = A_{0} P^{\alpha}$, which suggest $\alpha = 0.63$ and
$\alpha = 0.64$ for W1 and W2, respectively. } \label{mani04fig}
\end{figure}

Figure 2 presents a direct comparison of the transport
characteristics in W1 and U1 at $f = 50GHz$. Figure 2(a) and (b)
correspond to the experiment on the U-specimen  at $0.5K$, while
Fig. 2(c)  and (d) correspond to the W1 experiment at $T=1.5K$.
Here, the DPPH-ESR $B$-markers shown in Fig. 2(b) and 2(d) served
to calibrate $B$ for the corresponding experiments. Fig. 2(a) and
(c) show that both specimens exhibit Integer Fixed Points (IFP) at
the same $B^{-1}$. IFP's are points where the dark $R_{xx}$ trace
intersects the photo-excited $R_{xx}$ trace. IFP's occur in the
vicinity of $hf = n \hbar \omega_{c}$, where $\omega_{c}$ equals
the cyclotron frequency, and $n = 1,2,3...$.\cite{24} If two
material systems are similar, then one expects the IFP's to occur
at the same $B^{-1}$ at constant $f$, and this is what is observed
in the data. Thus, for example, the relative diffusion of Al into
the GaAs 2DES seems no different in the two systems, since greater
diffusion in one system than the other would lead to observable
differences in $m^{*}/m$ and thus the IFP's.

\begin{figure}[t]
\begin{center}
\leavevmode \epsfxsize=3 in
 \epsfbox {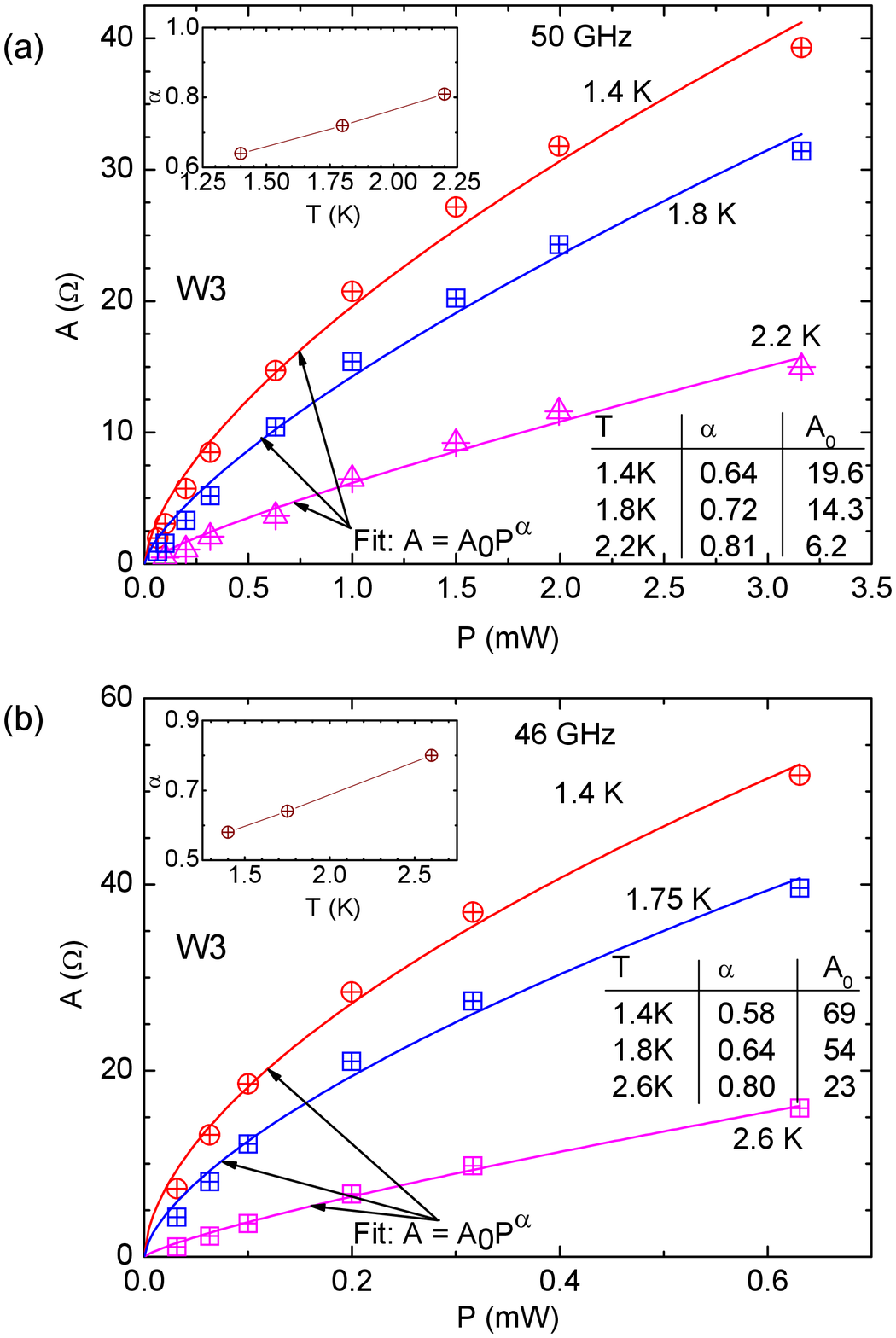}
\end{center}
\caption{(a) At $f=50GHz$, the amplitude, $A$, is plotted vs. $P$
for $T = 1.4K, 1.8K and 2.2K$ for specimen W3. Also shown are fits
to $A = A_{0} P^{\alpha}$. The fit-extracted $\alpha$ and $A_{0}$
are presented are presented in tabular form within the figure. The
inset shows the variation of $\alpha$ with $T$ at $50GHz$. (b) At
$f=46GHz$, the amplitude, $A$, is plotted vs. $P$ for $T = 1.4K,
1.75K and 2.6K$ for specimen W3. Also shown are fits to $A = A_{0}
P^{\alpha}$. The fit-extracted $\alpha$ and $A_{0}$ are presented
are presented in tabular form within the figure. The inset shows
the variation of $\alpha$ with $T$ at $46GHz$.} \label{mani04fig}
\end{figure}

\begin{figure}[t]
\begin{center}
\leavevmode \epsfxsize=3 in
 \epsfbox {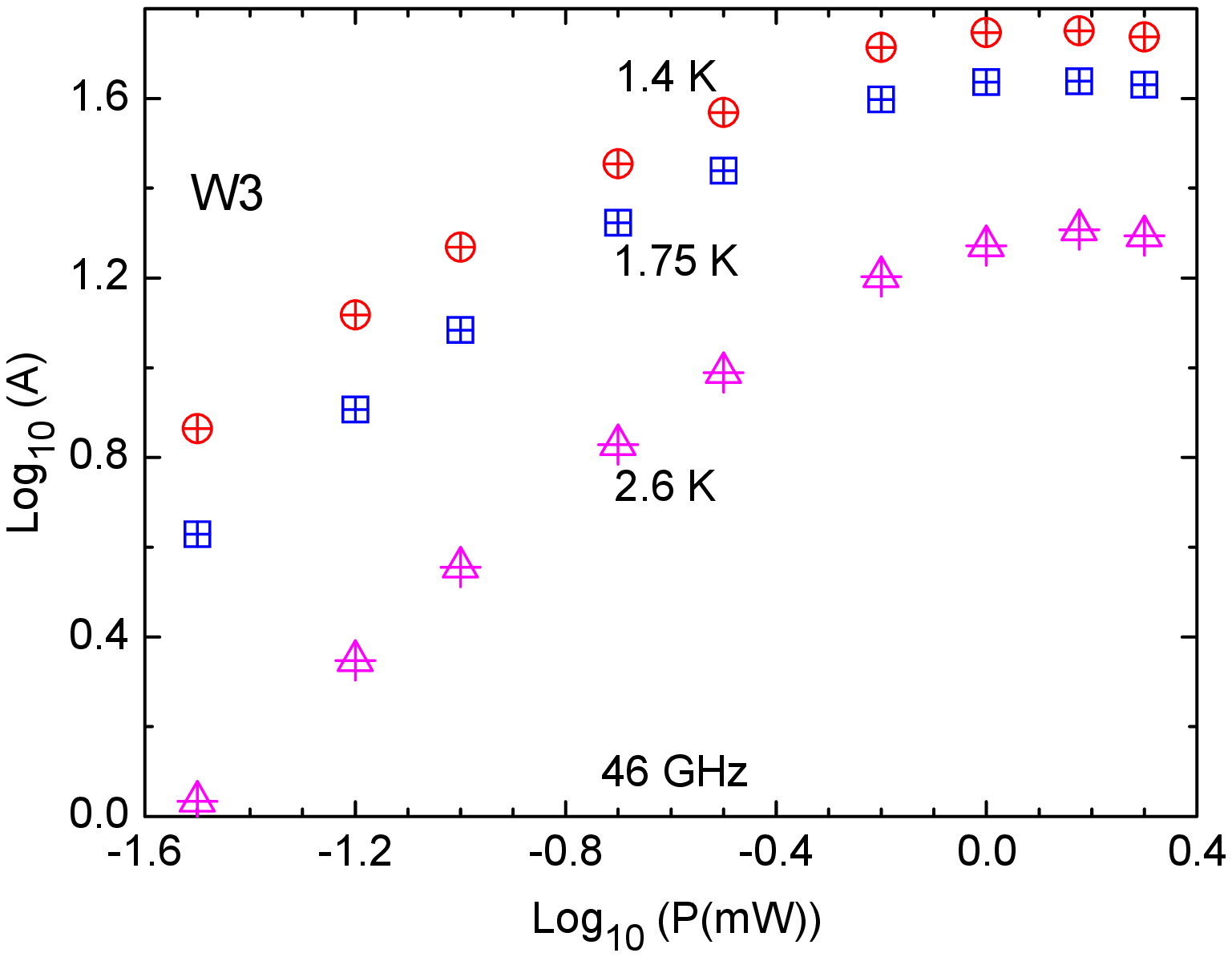}
\end{center}
\caption{This figure present a Log-Log plot of the amplitude $A$
of the exponentially-damped-sine-wave-fit (see text) vs. the
radiation-intensity $P$ at $46 GHz$ for specimen W3. At the
highest radiation intensities, the amplitude saturates and then
decreases with increasing $P$.} \label{mani04fig}
\end{figure}

To examine the growth of the radiation-induced oscillations with
$P$, Fig. 3 presents the $\Delta R_{xx}$ of U1 for several $P$ at
$50GHz$. Also shown in the figure are fits to the data using
$\Delta R_{xx}^{fit} = -A exp(-\lambda/B)sin(2 \pi F/B)$. Here, a
slowly varying background, approximately equaling the dark trace,
was removed from the photo-excited $R_{xx}$ data to realize
$\Delta R_{xx}$. Although this fit function includes three
parameters, $A$, $\lambda$, and $F$, the oscillation period in
$B^{-1}$ is independent of the radiation-intensity, and thus, $F$
is a constant. Further, the damping constant, $\lambda$,\cite{9}
turns out to be insensitive to $P$, as is evident in Fig. 3(b),
which shows $\lambda$ vs. $P$ for U1 and U2. The small difference
in $\lambda$ obtained for U1 and U2 in different cool-downs is
attributed to the PPC-effect dependence of the electronic
properties in the GaAs/AlGaAs system. Thus, the main free
parameter in the fit-function is the amplitude, $A$, of the
oscillations. In Fig. 3(c), we exhibit the fit extracted $A$ vs.
$P$ for U1 and U2. The figure shows the sub-linear growth of $A$
with $P$. Also shown are power law fits, $A = A_{0}P^{\alpha}$.
Here, $\alpha = 0.49$ and $\alpha = 0.50$ for U1 and U2,
respectively. $A_{0}$ varies between U1 and U2 because the
effective intensity attenuation factor is different in the two
experiments.

Fig. 4(a) shows power-dependent $\Delta R_{xx}$ vs $F/B$ at $50
GHz$ for W1, along with lineshape fits. As above, $F$ was
insensitive to $P$, and the $F$ of W1 agreed with the $F$ of U1,
see Fig. 2, and also U2. Fig. 4(b) exhibits $\lambda$ vs. $P$ for
W1 and W2. Again, $\lambda$ is insensitive $P$. Note that
$\lambda_{W} \approx 0.15 T < 0. 2 T \approx \lambda_{U}$. If the
damping is written as $exp(-\lambda/B) = exp
(-\pi/\omega_{c}\tau_{f}) \approx exp (-pT_{f}/B)$, where $T_{f}$
and $\tau_{f}$ represent finite frequency broadening temperature
and lifetime, respectively,\cite{9} then the $\lambda$ imply that
$T_{f,U} \approx 200 mK$ and $T_{f, W} \approx 150 mK$. Thus,
$T_{f,U} > T_{f,W}$, i.e., there is more broadening in the
U-specimen. Next, in Fig. 4(c), we exhibit the fit extracted $A$
vs. $P$ for W1 and W2. The figure suggests a nonlinear growth of
$A$ with $P$. Also shown are the fits, $A = A_{0}P^{\alpha}$,
which suggest that $\alpha = 0.63$ and $\alpha = 0.64$ for W1 and
W2, respectively.

For a third W-specimen labeled W3, Fig. 5 reports the influence of
the temperature on the growth of $A$ vs. $P$, where $A$ is
extracted, as before, from line-shape fits of the oscillatory data
to $\Delta R_{xx}^{fit} = -A exp(-\lambda/B)sin(2 \pi F/B)$.
Parenthetically, we note here that such fits suggested a general
insensitivity of $\lambda$ to the experimental parameters $f$,
$T$, and $P$. Indeed, $\lambda = 0.144 \pm 0.005$ served to fit
all the $f$, $T$, and $P$ covered in Fig. 5.

Fig. 5(a) and 5(b) indicate that, as expected, at a constant $P$,
$A$ grows with decreasing $T$ both at $f=50 GHz$ [Fig. 5(a)] and
$f=46GHz$ [Fig 5(b)]. Further, the figures show that the $A$ vs.
$P$ curves exhibit greater curvature at lower temperatures. The
$A$ vs. $P$ have been fit once again to $A = A_{0}P^{\alpha}$ in
order to quantify the observations; the fit-extracted $\alpha$ and
$A_{0}$ have been summarized in tabular form within Fig. 5(a) and
Fig. 5(b). These fit-extracted $\alpha$ have also been plotted vs.
$T$ in the inset to these figures. These insets suggest that
$\alpha$ decreases with decreasing temperatures, consistent with
the observed increased non-linearity at lower temperatures. Using
this result, one might attribute, at least in part, the smaller
$\alpha$ in the U-specimens, see Fig. 3(c), in comparison to the
W-specimens W1 and W2, see Fig. 4(c), to the lower temperature in
the U-measurements.

Note that, at $f=50GHz$, all three W-specimens, with comparable
material properties, exhibit the same $\alpha$, within
uncertainties, at the lowest pumped $^{4}$He temperatures, cf.
Fig. 4(c) and Fig. 5(a). In addition, a comparison of the $\alpha$
reported in Fig. 5(a) and Fig. 5(b) also suggests that reducing
the microwave frequency $f$ at a fixed $T$ tends to reduce the
$\alpha$, i.e., increase the non-linearity.

The nonlinear power law intensity variation reported here is
associated with modest excitation. It is known that the amplitude
of the radiation-induced magneto-resistance oscillations saturates
at higher $P$, and further, there is a "breakdown" and a decrease
in the amplitude at the highest $P$.\cite{13} The results shown
thus far in Fig. 1 - Fig. 5 do not correspond to such high $P$.

To illustrate the behavior in the higher $P$ regime, we exhibit in
Fig. 6,  the $A$ vs. $P$ at $f=46 GHz$, when efficient coupling
has been realized between the microwave source and the specimen,
i.e, when the sample has been located at the intensity peak in the
standing wave pattern realized within the waveguide sample-holder.
The $log_{10}-log_{10}$ plot exhibited in Fig. 6 shows $A$
saturation, and subsequent $A$ decrease with increasing $P$, at
the highest $P$, consistent with our previous report.\cite{13}

As mentioned, most theories have exhibited numerically evaluated
$R_{xx}$ or $\rho_{xx}$ including radiation-induced oscillations
at several $P$,\cite{24,31,33}, while Dmitriev et al,\cite{30}
have predicted that the amplitude of the radiation-induced
oscillations should increase linearly with $P$. The results
presented here suggest a nonlinear power law increase in a regime
characterized by modest excitation.

\section{Conclusion}

In summary, consistent experimental results have been obtained for
the U- and W- material so far as the essential features of the
radiation-induced magnetoresistance oscillations are concerned,
and this includes the "1/4-cycle" phase shift.\cite{8} In
addition, nonlinear growth is observed in the amplitude of
radiation-induced magneto-resistance oscillations with $P$ in both
materials. Although fits have been utilized here to quantify the
non-linear intensity dependence, such fits are not necessary to
perceive the basic non-linear intensity dependence in the raw
data.

\section{Acknowledgements}

R.G.M. is supported by the Army Research Office under
W911NF-07-01-0158 and the DOE under DE-SC0001762.


\begin{thebibliography}{19}
\bibitem{1} M. Tinkham, Introduction to Superconductivity, 2nd.
ed. (McGraw-Hill, New York, 1996).

\bibitem{2} R. E. Prange and S. M. Girvin, The Quantum Hall
Effect, 2nd. ed. (Springer, New York, 1990).

\bibitem {3} S. Das Sarma and A. Pinczuk, Perspectives in Quantum Hall
Effects (Wiley, New York, 1996).

\bibitem{4} R. G. Mani, J. H. Smet, K. von Klitzing, V. Narayanamurti,
W. B. Johnson, and V. Umansky, Nature(London) \textbf{420}, 646
(2002).

\bibitem{5} M. A. Zudov, R. R. Du, L. N. Pfeiffer, and K. W. West, Phys.
Rev. Lett. \textbf{90}, 046807 (2003).

\bibitem{6} S. I. Dorozhkin, JETP Lett. \textbf{77}, 577 (2003).

\bibitem{7}R. G. Mani, V. Narayanamurti, K. von Klitzing, J. H. Smet, W. B.
Johnson, and V. Umansky, Phys. Rev. B\textbf{69}, 161306 (2004);
Phys. Rev. B\textbf{70}, 153310 (2004).

\bibitem{8}R. G. Mani et al., Phys. Rev. Lett. \textbf{92}, 146801
(2004).

\bibitem{9} R. G. Mani et al., Phys. Rev. B\textbf{69}, 193304
(2004).

\bibitem{10} S. A. Studenikin et al., Sol. St. Comm. \textbf{129}, 341
(2004).

\bibitem{11} R. R. Du et al., Physica E (Amsterdam) 22, 7, (2004).

\bibitem{12} R. L. Willett, L. N. Pfeiffer, and K. W. West, Phys.
Rev. Lett. \textbf{93}, 026604 (2004).

\bibitem{13}R. G. Mani, Physica E (Amsterdam) \textbf{22}, 1 (2004);
\textit{ibid.} \textbf{25}, 189 (2004).

\bibitem{14}R. G. Mani, Appl. Phys. Lett. \textbf{85}, 4962 (2004);
IEEE Trans. on Nanotech. \textbf{4}, 27 (2005); Phys. Rev. B
\textbf{72}, 075327 (2005); Appl. Phys. Lett. \textbf{91}, 132103
(2007); Sol. St. Comm. \textbf{144}, 409 (2007); Appl. Phys. Lett.
\textbf{92}, 102107 (2008); Physica E, \textbf{40}, 1178 (2008).

\bibitem{15} B. Simovic et al., Phys. Rev. B\textbf{71}, 233303
(2005).

\bibitem{16} J. H. Smet et al., Phys. Rev. Lett. 95, 118604 (2005).

\bibitem{17} Z. Q. Yuan et al., Phys. Rev. B\textbf{74},
075313 (2006).

\bibitem{18} S. A. Studenikin et al., Phys. Rev. B\textbf{76}, 165321 (2007).

\bibitem{19} K. Stone et al., Phys. Rev. B\textbf{76}, 153306 (2007).


\bibitem{20} A. Wirthmann et al., Phys. Rev. B\textbf{76}, 195315 (2007).

\bibitem {21} S. Wiedman et al., Phys. Rev. B 78, 121301(R)
(2008).

\bibitem{22} A. T. Hatke, M. A. Zudov, L. N.
Pfeiffer, and K. W. West, Phys. Rev. Lett. \textbf{102}, 086808
(2009).

\bibitem{23} R. G. Mani et al., Phys. Rev. B\textbf{79}, 205320
(2009).

\bibitem{24} A. C. Durst, S. Sachdev, N. Read, and S.
M. Girvin, Phys. Rev. Lett. \textbf{91}, 086803 (2003).

\bibitem{25}A. V. Andreev, I. L. Aleiner, and A. J. Millis, Phys. Rev. Lett.
\textbf{91}, 056803 (2003).

\bibitem{26}V. Ryzhii and A. Satou, J. Phys. Soc. Jpn. \textbf{72}, 2718
(2003).

\bibitem{27}X. L. Lei and S. Y. Liu, Phys. Rev. Lett.
\textbf{91}, 226805 (2003).

\bibitem{28} P. H. Rivera and P. A. Schulz, Phys. Rev. B\textbf{70},
075314 (2004); P. H. Rivera, A. L. C. Pereira, and P. A. Schulz,
Phys. Rev. B 79, 205406 (2009)

\bibitem{29} S. A. Mikhailov, Phys. Rev. B\textbf{70}, 165311 (2004).

\bibitem{30}I. A. Dmitriev et al., Phys. Rev. B\textbf{71}, 115316 (2005).

\bibitem{31}J. Inarrea and G. Platero, Phys. Rev. Lett. \textbf{94}, 016806
(2005).

\bibitem{32} A. Auerbach, I. Finkler, B. I. Halperin, and A.
Yacoby, Phys. Rev. Lett. \textbf{94}, 196801 (2005).

\bibitem{33} X. L. Lei and S. Y. Liu, Phys. Rev. B\textbf{72}, 075345
(2005).

\bibitem{34} A. D. Chepelianskii, A. S. Pikovsky and D. L.
Shepelyansky, Eur. Phys. J. B\textbf{60}, 225 (2007).

\bibitem{35} A. Auerbach and G. V. Pai, Phys. Rev. B\textbf{76}, 205318
(2007).

\bibitem{36} I. A. Dmitriev, A. D. Mirlin, and D. G. Polyakov, Phys.
Rev. B\textbf{75}, 245320 (2007).

\bibitem{37}J. Inarrea and G. Platero, Phys. Rev. B\textbf{78}, 193310 (2008).

\bibitem {38} M. Khodas and M. Vavilov, Phys. Rev. B\textbf{78}, 245319
(2009).

\bibitem{39} I. Finkler and B. I. Halperin, Phys. Rev. B\textbf{79},
085315 (2009).

\bibitem{40} A. D. Chepelianskii and D. L. Shepelyansky, Phys.
Rev. B \textbf{80}, 241308 (2009).

\bibitem{41} D. Konstantinov and K. Kono, arxiv:0910.3040

\end{thebibliography}
\pagebreak

\end{document}